\documentclass[
twocolumn,
]{ceurart}

\usepackage{listings}
\begin{document}

\copyrightyear{2023}
\copyrightclause{Copyright for this paper by its authors.
  Use permitted under Creative Commons License Attribution 4.0
  International (CC BY 4.0).}

\conference{VLDB 2023 PhD Workshop, co-located with the 49th International Conference on Very Large Data Bases (VLDB 2023), August 28, 2023, Vancouver, Canada}

\title{Improving Data Minimization through Decentralized Data Architectures}

\author{Ilaria Battiston}[%
orcid=0009-0007-6180-364X,
email=ilaria@cwi.nl,
]
\address{supervised by Peter Boncz}
\address{Centrum Wiskunde \& Informatica, Amsterdam, The Netherlands}


\begin{abstract}
In this research project, we investigate an alternative to the standard cloud-centralized data architecture. Specifically, we aim to leave part of the application data under the control of the individual data owners in 
decentralized personal data stores. Our primary goal is to increase data minimization, i.\:e., enabling more sensitive personal data to be under the control of its owners while providing a straightforward and efficient framework to design architectures that allow applications to run and data to be analyzed. To serve this purpose, the centralized part of the schema contains aggregating views over this decentralized data. 
We propose to design a declarative language that extends SQL, for architects to specify different kinds of tables and views at the schema level, along with sensitive columns and their minimum granularity level of their aggregations. Local updates need to be reflected in the centralized views while ensuring privacy throughout intermediate calculations; for this we pursue the integration of distributed materialized view maintenance and multi-party computation (MPC) techniques. 
We finally aim to implement this system, where the personal data stores could either live in mobile devices or encrypted cloud storage, in order to evaluate its performance properties. 
\end{abstract}


\maketitle

\section{Introduction}
Organizations almost invariably adopt a data architecture based on centralization for their IT systems, retaining information in analytical stores. 
Making an organization data-driven, i.\:e. relying more on data science and machine learning for decision-making, is a driver for expanding data architectures, strengthening the trend of widespread collection of personal data through centralized and typically cloud-based systems. 
Such systems facilitate data processing; however, they present concerns pertaining to security and confidentiality. The organization is in control of all the user data,  causing owners, often private citizens, to lose control over this sensitive data. Full centralisation of detail-level sensitive data increases the exposure of the organization to ransomware attacks - as well as the needed cloud resources.

In the PhD research plan outlined in this paper, we investigate partially decentralized alternatives to the fully centralized state of affairs, to give users more control over their personal data and reduce processing costs and risks for organizations running applications. We thus aim for a generic infrastructure that pushes the boundaries of \textit{data minimization}~\cite{43-Pfitzmann09aterminology}, i.\:e., leaving more data under the control of its creators/owners while still allowing easy construction of centralized applications, leveraging relational database systems and their declarative properties. 

This paper is structured as follows: Sections 2 and 3 motivate the need for responsible decentralized architectures, along with open questions to be foreseen based on the proposed system. Section 4 illustrates the research background and previous approaches. Section 5 describes the envisioned architecture, and we conclude and discuss the next steps in Section 6. 

\section{Motivation}

\textit{Data minimization} is the principle of limiting the data that an organization collects to the minimum needed for the purpose~\cite{43-Pfitzmann09aterminology}, and is a crucial concept in privacy regulations such as the GDPR.  

Still, digital services often {\em do} need private information, so data minimization in itself does not avoid sensitive user data being collected. However, what is necessary for a purely centralized data architecture might not be needed in the partially decentralized data architectures for which we aim to develop a generic infrastructure.

A possible use case considers fitness tracker applications, whose popularity has raised concerns regarding the privacy of collected personal data. For example, this can include user profile information, activity statistics, health metrics, and geographical coordinates. We argue that for showing the top-10 runners in a circuit or the distribution of running times, it is not necessary to bring all detail data to the central application database. Our approach allows to decentralize sensitive details - such as health metrics and the coordinates of runs - in {\em personal data stores}, and only transmit aggregated running data to the central application database. A personal data store is a personal database purely under the user's control, e.g., a local database kept on a personal device, and/or stored in a separate cloud service; but encrypted with a personal key that only the user holds.

This 
project aims to {\em reduce} sensitive user data storage in central databases without compromising user privacy, providing future application architects with a simple solution without impacting their ability to create compelling applications, while giving end users more control over their personal data. Our research may also contribute to advances in differential privacy.


We focus on the design of a declarative framework in which information architects can use SQL (i.\:e. relational database technology to leverage the analytical properties of structured data) to split a data management architecture between a {\em centralized} and {\em decentralized} part, as well as on a secure implementation of this framework in a real system, and an evaluation of its efficiency properties. This concept gives users ownership and cryptographic security over their {\em personal data stores}, allowing an end-user inspection of the aggregated queries ordered by the central database. 
A major research question concerns privacy-preserving mechanisms for incremental materialized view maintenance~\cite{Gupta1999MaintenanceOM} in this setting: we want to hide from the central database what each user's individual contribution to a sensitive materialized aggregate is.

As a platform for RDDA prototyping, we choose the open-source novel data management system DuckDB~\cite{48-DBLP:conf/sigmod/RaasveldtM19}, offering the ability to run analytical queries efficiently even on low-power devices. However, since our framework strongly relies on SQL, it will be easily portable.

\section{Research Questions}
The concepts proposed so far raise a number of research questions:
\begin{enumerate}
\item How can we specify a decentralized data architecture in a declarative language (e.\:g. a SQL extension), and what properties or constraints should it contain?
\item How can we apply cryptographic techniques to incrementally maintain materialized views in a manner that does not leak more than understandable constraints stipulate, controlling the amount of privacy leakage?
\item How can we help establish trust from end users, providing insight and control over personal data and attesting that the service or application implementing the framework plays by its rules?
\end{enumerate}

\section{Related Work}
The best-known attempt at creating a unified personal data store is Tim Berners Lee's SOLID project\footnote{https://solidproject.org/}, but this does not envision central analytical queries.
We think that giving centralized query services access to analytics over whole fleets of personal datastores, via privacy-controlled materialized views, will provide extra value that may help adoption of the concept of personal data stores, which until now has been lackluster.

Prior query-oriented decentralized infrastructures are mainly found in specific applications, such as real-time cellular network analytics systems exploiting geo-partitioning of input data~\cite{31-DBLP:conf/nsdi/IyerLS15}. Distributed and federated query processing are also the subject of extensive research. However, these always assume a free choice in data placement decisions~\cite{10-DBLP:journals/corr/abs-1805-08520, 41-DBLP:phd/dnb/Muhleisen12}, rather than ascertaining that personal data is kept private and under the control of the end-user. 
Federated query processing systems also assume an online mode of operation.
Our assumption that only the end users can access their personal data stores leads to an approach where updates from the user side must trickle to a central query-answering facility later, which in turn leads to incremental view maintenance (IVM).
IVM has been studied extensively~\cite{ahmad2012dbtoaster, budiu2022dbsp}, and we aim to build on this work.
However, an additional complication is that the incremental maintenance actions may leak information.
Therefore, we study mechanisms concerning data processing and supporting cryptographic methods in which updates of multiple users are combined to form more coarse-grained materialized view updates.

A related research field to protect sensitive data is differential privacy~\cite{25-DBLP:conf/tamc/Dwork08}, the technique of adding noise to the data to obscure any individual's information while maintaining the statistical accuracy of the overall set. 
We think that our decentralized data architectures are a potential use case to build new forms of {\em decentralized} differential privacy, and note that research in this area tends to assume a central database.
We also note that state-of-art differential privacy approaches within database systems such as Pinq~\cite{mcsherry2009privacy} and Google DP\footnote{https://github.com/google/differential-privacy} work best with aggregated data, matching our concept of aggregating materialized views.
Finally, there is a relevant time dimension: (i) {\em recency} of IVM updates will be part of our trade-offs against violating privacy constraints, and (ii) data architects may want to limit the lifetime of data in the materialized views.
Therefore, we plan to incorporate specific stream processing elements in our framework.
The most relevant research we perceive is the Dataflow model~\cite{13-DBLP:journals/pvldb/AkidauBCCFLMMPS15}, which was the first to clearly separate the concepts of data arrival time and event time in stream processing.
We think these notions will be necessary to define accuracy metrics of our materialized views.

\section{Architecture}
\begin{figure}
  \centering
  \includegraphics[width=\linewidth]{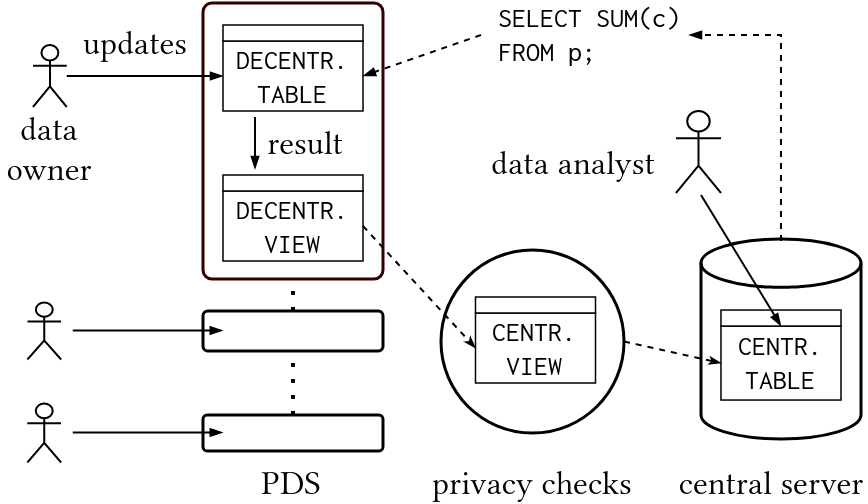}
  \caption{The proposed architecture. Decentralized views collect deltas computed from updates, and send aggregated query results to the central server. Before permanently storing the data, privacy checks are performed to assess whether anonymity can be granted.
  }
  \label{fig:rdda}
\end{figure}

\autoref{fig:rdda} shows our decentralized architecture. The infrastructure includes three components: in the first (left), multiple users query and update their personal data stores containing only their private data.
The personal data stores reside in encrypted cloud-based storage, whose key belongs to the individual user, and updates are reflected there. The second component (middle) is a secure analytics infrastructure responsible for applying deltas and checking whether the collected data respects privacy constraints without leaking information. The third component (right) is a centralized database.

Organizations can host online applications using such central databases (as is standard now), as well as run analytical workloads, with the difference that some of this data is stored in materialized views of which the detailed underlying data stems from personal data stores, left under user control. Periodically, upserts from aggregate queries are applied to central materialized views. To maintain privacy guarantees, we establish a minimum granularity level, checking whether each group contains a sufficient number of elements. Privacy rules can be user or system defined: how to choose appropriate bounds is still an open research question. However, this step must be performed without revealing any content before guaranteeing that information is privacy-preserving: a possible technique is {\em multi-party computation}~\cite{10.1145/3387108}.

Our infrastructure then relays these results to the central server, where data analysts can query centralized tables.
The framework also periodically transmits central updates to the replicated tables, and general statistics on the completeness of the incrementally maintained views are available to the central database in order to give accuracy bounds on query processing. When individual users of an application, on the other hand, need query processing on their personal data stores or the replicated tables, this involves only the first tier and can be done locally.

All the components will be specified in a declarative manner: application architects are unlikely to be experts in privacy-conscious decentralized data management. We, therefore, propose our language to be an extension of SQL, hoping to expedite the adoption of our framework. 

\textit{Decentralized tables} represent personal data stores and can only implement references to other local tables. These can be seen as horizontal partitions (row groups), assigned an implicit identifier at the moment of database initialization. Operations can therefore be executed concurrently on different partitions, allowing for improved performance and smaller transaction scope, similar to the Google Spanner architecture~\cite{39966} but only requiring explicit declaration during the table creation process.

\textit{Centralized tables}, on the other hand, are under the control of the application architect and formally represent the union of aggregations over the partitions. \textit{Replicated tables} can also be defined, containing overviews to be periodically propagated from the second to the first tier, such as public dashboards.

We introduce an additional concept of \textit{decentralized views}, 
defined over the tables in the personal data store to identify those pieces of data that may be exported centrally. This additional abstraction layer is intended to give end users more insight into what
data is centrally readable. Decentralized views contain the records to be communicated to the central entity, which are then stored in \textit{centralized views}. 

In addition, centralized views may introduce \textit{time windows}, either in terms of logical (event) time from the data or actual update time. Centralized views can therefore be defined to retain only data from a limited number of such windows. The purpose is to aid information architects in realizing retention limits directly through a SQL specification. 
This feature also broadens our research question 
toward incremental streaming view maintenance.

The column specifications in our table and view definitions will allow to  e.\:g. add  randomized noise to facilitate building differential privacy on top of our framework.
They also allow defining \textit{sensitive} columns, as well as \textit{minimum aggregation granularity}, such as a minimum of at least e.\:g., 100 values for an aggregate result tuple that involves this sensitive column to be included in it.
We aim to develop declarative rules to identify potentially privacy-breaking queries, which may necessitate SQL extensions to provide an additional layer of security.

The previously described design provides an easy way for application architects to specify the components of our infrastructure, however, it alone does not guarantee to protect data owners from possible malevolence. Aggregate data could still contain sensitive information or not have a sufficient level of granularity, failing to provide anonymization. For example, it would be easy to recognize individuals belonging to a group with only one element. However, such calculations cannot be performed until information from multiple PDS is obtained.

In order to mitigate the potential risk of unauthorized access to database records, the offloading of computations to a third-party entity can be considered. Nonetheless, it is crucial to establish a foundation of trust in these additional service providers. A possible solution is to employ 3-way multi-party computation (MPC), either with different cloud providers or a peer-to-peer system, to hide information while it is being processed until it respects our privacy constraints. The state-of-art system Secrecy\cite{DBLP:journals/corr/abs-2102-01048} allows secure collaborative analytics through oblivious SQL queries. We plan to extend this framework with IVM techniques to be able to perform bulk updates and insertions.

Expensive IVM operations such as joins can be performed locally on PDS in plain text; results are then applied to decentralized views and sent over a secure communication channel (TLS) to be ultimately stored in centralized views. Our MPC servers, therefore, only need to append new rows or update the aggregated values in single tables, which can be performed through cheap oblivious arithmetic operations.

Establishing trust in the organization responsible for setting up our infrastructure, ensuring they fulfill their claims, remains a prerequisite in this methodology. This can be achieved through various approaches, including enabling transparency by exposing all server traffic and resource utilization, conducting audits, and leveraging the use of open source technologies. However, our exploration of efficient encryption within incremental view maintenance is ongoing, and we remain open to additional ideas that could enhance our approach.

\section{Conclusion and Future Work}
In this paper, we outlined a PhD research agenda for the RDDA project, aiming to design and implement a declarative framework for partially {\em decentralized} data architectures, that pushes the boundaries of {\em data minimization} by leaving data in personal data stores under the control of the end users while still enabling application architects to perform centralized data analyses on top of them. 
We outlined the most important questions that we will research and provided a short overview of the SQL-defined components of this framework, along with initial ideas on leveraging MPC to provide privacy-preserving incremental view maintenance.
We aim to release an initial open source implementation of this framework, based on DuckDB. Furthermore, we plan to investigate the qualitative and quantitative properties to evaluate our infrastructure in order to assess its usability and performance.

\bibliography{sample-ceur}

\end{document}